\documentclass[%
reprint,
amsmath,amssymb,
prmaterials,
floatfix,
]{revtex4-2}

\usepackage{tikz}

\usepackage{xr-hyper}
\makeatletter
\newcommand*{\addFileDependency}[1]{
	\typeout{(#1)}
	\@addtofilelist{#1}
	\IfFileExists{#1}{}{\typeout{No file #1.}}
}
\makeatother

\newcommand*{\myexternaldocument}[1]{%
	\externaldocument{#1}%
	\addFileDependency{#1.tex}%
	\addFileDependency{#1.aux}%
}
\myexternaldocument{Supplemental} 
\usepackage{hyperref}

\usepackage{graphicx}
\usepackage{xcolor}
\usepackage{dcolumn}
\usepackage{bm}

\begin{document}
	
	\preprint{APS/123-QED}
	
	\title{Benchmarking of different strategies to include anisotropy in a curvature-driven multi-phase-field model}
	
	\author{Martin Minar}
	\email{martin.minar1132@gmail.com}
	\author{Nele Moelans}%
	\affiliation{KU Leuven, Department of Materials Engineering}
	
	\date{\today}
	
	\begin{abstract}
		Two benchmark problems for quantitative assessment of anisotropic curvature driving force in phase field method were developed and introduced. Both benchmarks contained an anisotropically shrinking grain in homogeneous matrix. The first benchmark was a shrinking Wulff shape and in the second, such inclination dependence of the kinetic coefficient was added so that the shrinkage was isotropic. In both cases the match to the expected shape was quantified by means of Hausdorff distance and the shrinkage rate was analytically expressed. Three different ways of interface energy anisotropy inclusion in a multi-phase field model were compared. Because their performance was comparable, they were tested in an additional benchmark problem, which concerned direct measurement of equilibrium triple junction angles. Based on this benchmark, only one of the three strategies to include anisotropy was reliable in strongly anisotropic systems.
	\end{abstract}
	
	\maketitle
	
	
	\section{Introduction}
	\label{sec_Intro}
	
	In order to confidently and honestly interpret results of quantitative phase field simulations, both physical models and numerical implementations must be validated and verified~\cite{Jokisaari2017}. Recent initiative PFHub addresses the need for benchmarking of the multitude of software and numeric approaches for solving the phase field governing equations. However, benchmarks validating anisotropic interface energy or multi phase field models have not been included yet.
	
	Grain coarsening is assumed to be curvature-driven, hence it is an appealing application for only-interface-driven multi-phase field models. Real interfaces (including the grain boundaries) exhibit anisotropic (inclination-dependent) interface energies ~\cite{Olmsted2009,Bulatov2014}, hence a significant effort was made to introduce this feature in some multi-phase field models too~\cite{Garcke1999,Toth2015,Salama2020,Kazaryan2000,Wendler2011}.
	
	The models have different formulations, but the differences in their behavior are not obvious, especially because there is no unified validation procedure for quantitative comparison. Systematic and reproducible parametric studies in well-defined problems are needed for quantitative assessment of phase field models reliability. 
	
	There are several ways to introduce anisotropic interface energy in the Allen-Cahn equation (irrespective of whether the model is single- or multi-phase field). Tschukin \cite{Tschukin2017} proposes the terminology of \textit{classical} and \textit{natural} models (corresponding to the notation VIW and VIE, respectively, as used by Fleck~\cite{Fleck2011}). The classical models~\cite{Kobayashi1993,McFadden1993,Taylor1998,Eggleston2001,Wheeler2006}, introduce anisotropy in the gradient energy coefficient, whereas the natural models~\cite{Ma2006,Torabi2009,Fleck2011,Moelans2008} in both the gradient energy coefficient and homogeneous energy density barrier. The main difference is that in the first case the diffuse interface width varies proportionally to the local interface energy, whereas in the latter they are decoupled and the interface width is constant. Another notable approach uses Finsler geometry to replace the Euclidean metric in the simulation domain by an anisotropic one~\cite{Bellettini1996,Benes2003}.
	
	Usually, the validation of models with inclination-dependent interface energy was carried out by visual comparison of the phase field contours to corresponding Wulff shapes for single or several values of strength of anisotropy~\cite{Garcke1999,Eggleston2001,Fleck2011,Ma2006,Tschukin2017}. Such approach does not reveal the limits of reliability of these models, though.
	
	This paper proposes two benchmarks for quantitative assessment of a) the anisotropic curvature driving force, and b) the anisotropic curvature driving force in combination with anisotropic kinetic coefficient. Both are 2-phase systems, hence could be simulated using a single phase field model. Nevertheless, because the multi-phase field models allow simulations of multiple phases, they have wider application potential than single phase models. For this reason, the validations are demonstrated on a multi-phase field model~\cite{Moelans2008}. However, the nature of these benchmarks is independent on the model formulation. 
	
	In order to provide more complete comparison of the models, another supplementary benchmark was carried out, which determined equilibrium triple junction angles.
	
	The multi-phase field model by Moelans~\cite{Moelans2008} is an established~\cite{Miyoshi2020} quantitative phase field model of grain growth with anisotoropic grain boundary properties. Using asymptotic analysis, Moelans derived that in the original model~\cite{Moelans2008}, the local interface energy and width are related to three model parameters. Because these are two equations of three variables, the system is undetermined and one of the model parameters is free. This degree of freedom introduces the possibility of many different parameters assignment strategies, all of which represent the same physical input. The effect of different but equivalent parameters choices can thus be investigated in this model. Moelans proposed such parameters assignment strategy~\cite{Moelans2008}, which assured constant interface width irrespective of the strength of anisotropy in interface energy (a kind of \textit{natural} formulation). Two of the three model parameters ($\gamma$ and $\kappa$) were made anisotropic in order to achieve such behavior. However, this approach (here denoted IWc - \textit{Interface Width constant}) does not reproduce well the angles between interfaces in triple junctions for stronger anisotropies. This was already first noted by Moelans in~\cite{Moelans2010}. An alternative parameters assignment strategy with only parameter $\gamma$ anisotropic was used in~\cite{Ravash2017,Miyoshi2020}, but no systematic comparison was made. In this approach, the interface width is not constant in anisotropic systems, but it is not simply classifiable as \textit{classical} anisotropy formulation, because the gradient energy coefficient is constant. It will be denoted IWvG (\textit{Interface Width variable and Gamma anisotropic}). The third compared parameters assignment strategy is the \textit{classical} formulation, varying only gradient energy coefficient $\kappa$ to achieve the desired interface energy anisotropy (denoted IWvK - \textit{Interface Width variable and Kappa anisotropic}). Additionally, the inclination dependence of interface energy in IWvG and IWvK have not yet been addressed in the framework of Moelans' model. 
	
	This paper is organized as follows: firstly, the base model and its three variants are introduced, including the inclination dependence in interface energy. Secondly, the methodology is explained in detail, which involves quantitative matching of the shrinking shape to the analytic one and also the determination of shrinkage rate (also known analytically). The approach taken in triple junction angles determination is explained as well. Then, the validations were carried out in the order: shrinking Wulff shape, kinetically compensated anisotropically shrinking circle and triple junction angles. For all simulations the effect of interface width and number of grid points through the interface are investigated by re-running the simulations using different numerical settings.
	
	\section{Phase field model}
	The system consists of $n$ non-conserved continuous-field variables (further denoted phase fields) $\eta_1(\mathbf{r},t), \eta_2(\mathbf{r},t),\dots,\eta_n(\mathbf{r},t)$, which are functions of space and time. The total free energy of the system is expressed as functional of the phase fields and their gradients $\nabla\eta_1(\mathbf{r},t), \nabla\eta_2(\mathbf{r},t),\dots,\nabla\eta_n(\mathbf{r},t)$
	\begin{equation} \label{eq_def_totalE_ver1}
		F = \int_V \Bigg\{ m f_0(\vec{\eta}) + \frac{\kappa}{2}\sum_{i=1}^n(\nabla \eta_i)^2 \Bigg\} \mathrm{d}V \,,
	\end{equation}
	where the homogeneous free energy density $f_0(\vec{\eta}) = f_0(\eta_1, \eta_2,\dots,\eta_n)$ is expressed as 
	\begin{equation}
		f_0(\vec{\eta}) = \sum_{i=1}^n\left(\frac{\eta_i^4}{4} - \frac{\eta_i^2}{2} \right) +\gamma\sum_{i=1}^n\sum_{i>j}\eta_i^2\eta_j^2 + \frac{1}{4} \,.
	\end{equation}
	The parameters $m,\kappa, \gamma$ are model parameters, which together define interface energy and interface width (see the following section for more details).
	
	The governing equations for each phase field $\eta_p$ are obtained based on the  functional derivative of the free energy functional with respect to $\eta_p$, assuming that the phase-fields are non-conserved, i.e. 
	\begin{equation}
		\label{eq_ACeq_governing}
		\frac{\partial \eta_p}{\partial t} = -L\frac{\delta F}{\delta \eta_p} = -L \left[ \frac{\partial f}{\partial \eta_p} - \nabla\cdot\frac{\partial f}{\partial(\nabla \eta_p)} \right] \,,
	\end{equation}
	where $L$ is the kinetic coefficient (also dependent on the model parameters), $f$ is the full integrand in~\eqref{eq_def_totalE_ver1} and $\nabla\cdot\partial f/\partial(\nabla \eta_p)$ is divergence of vector field $\partial f/\partial(\nabla \eta_p)$ defined by relation
	\begin{equation}
		\frac{\partial f}{\partial(\nabla \eta_p)} = \frac{\partial f}{\partial(\partial_x\eta_p)}\mathbf{n}_x + \frac{\partial f}{\partial(\partial_y \eta_p)}\mathbf{n}_y + \frac{\partial f}{\partial(\partial_z \eta_p)}\mathbf{n}_z
	\end{equation}
	with $\partial_x,\partial_y,\partial_z$ being operators for unidirectional derivatives in the corresponding directions and $\mathbf{n}_x,\mathbf{n}_y,\mathbf{n}_z$ coordinate base vectors. 
	
	\subsection{Isotropic model}
	\label{sec_Models}
	In a system with uniform grain boundary properties, the interface energy is equal for all interfaces and hence the phase-field model parameters $m,\kappa, \gamma$ (and interface width $l$) are constant in the system.
	
	Then, using expression~\eqref{eq_ACeq_governing}, the governing equation for each phase field $\eta_p$ takes the following form 
	\begin{equation}
		\frac{\partial \eta_p}{\partial t} = -L\left[ m\left( \eta_p^3-\eta_p +  2\gamma\eta_p\sum_{j\neq p}\eta_j^2 \right) - \kappa\nabla^2\eta_p \right] 
	\end{equation}
	
	The interface energy $\sigma$ of the system is related to the model parameters via 
	\begin{equation}\label{eq_IE}
		\sigma = g(\gamma)\sqrt{m\kappa} \,,
	\end{equation}
	where $g(\gamma)$ is a non-analytic function of parameter $\gamma$. The interface width $l$ is expressed as
	\begin{equation} \label{eq_IW}
		l = \sqrt{\frac{\kappa}{mf_{0c}(\gamma)}} \,,
	\end{equation}
	where $f_{0c}(\gamma)$ is the value of $f_0(\eta_{i,cross},\eta_{j,cross})$ in the points where the two phase fields $\eta_i,\eta_j$ cross. $f_{0c}(\gamma)$ is a non-analytic function too. Values of both $g(\gamma)$ and $f_{0c}(\gamma)$ were tabulated and are available in~\cite{Ravash2017}. Both functions are positive and monotonously rising.
	
	Usually, the interface energy $\sigma$ is known as material property and $l$ is chosen for computational convenience, together with $\gamma$. Then, the parameter values are assigned from the following formulae
	
	\begin{equation}\label{eq_def_kappa}
		\kappa = \sigma l\frac{\sqrt{f_{0c}(\gamma)}}{g(\gamma)} \approx \frac{3}{4}\sigma l
	\end{equation}
	\begin{equation} \label{eq_def_m}
		m = \frac{\sigma}{l}\frac{1}{g(\gamma)\sqrt{f_{0c}(\gamma)}} \approx 6 \frac{\sigma}{l}
	\end{equation}
	\begin{equation}\label{eq_def_L}
		L = \frac{\mu}{l}\frac{g(\gamma)}{\sqrt{f_{0c}(\gamma)}} \approx \frac{4}{3} \frac{\mu}{l}
	\end{equation}
	The symbol $\mu$ stands for interface mobility. The approximate relations above hold exactly when $\gamma=1.5$ and are well applicable when $0.9 \leq \gamma \leq 2.65$ ~\cite{Moelans2008}. 
	
	\subsection{Anisotropic model and parameters assignment strategies}
	Two cases of interface energy anisotropy may occur, together or separately. Firstly, in the system there may be multiple interfaces with different interface energies (termed \textit{misorientation dependence} in~\cite{Moelans2008}, here \textit{pair-wise isotropy} for greater generality). Secondly, there may be an interface with inclination-dependent interface energy. Additionally, the kinetic coefficient $L$ can be inclination dependent. 
	
	In both cases of anisotropy in interface energy, some of the model parameters $m,\kappa, \gamma$ must become spatially dependent in order to assure correct local representation of the interface energy and width. In other words, the equations~\eqref{eq_IE}~and~\eqref{eq_IW} must hold in every point of the anisotropic system.
	These two equations locally form an undetermined system of three variables, hence one of the model parameters is free and many different parameters assignment strategies are possible.
	
	In this paper, the parameter $m$ is always a constant, because when $m$ was spatially varied~\cite{Moelans2008}, the model behavior in multijunctions was reported to be strongly affected by the interface width. Such model would be non-quantitative and thus will not be further regarded in this paper.
	
	Three different parameters assignment strategies are considered, which differ in value of $m$ and further in which of parameters $\kappa,\gamma$ is constant and which varies in space to keep equation~\eqref{eq_IE} valid. The three strategies are denoted: IWc (variable $\gamma,\kappa$ so that interface width is constant  \cite{Moelans2008}), IWvG (variable interface width and $\gamma$ \cite{Ravash2017}) and IWvK (variable interface width and $\kappa$).   Table~\ref{tab_models_comparison} summarizes, which parameters are kept constant and which vary to capture the anisotropy in the different strategies. The detailed procedure of the parameters assignment and ways to control the width of the narrowest interface are described in~S.I of the Supplemental Material~\cite{Minar2021suppl}. Note that for the IWc model we propose a single-step parameters determination procedure, which is more predictable and simpler than the original iterative one~\cite{Moelans2008}. The two are equivalent, though.\\
	Below follow details about the incorporation of pair-wise isotropy and inclination-dependence in the model. 
	
	\begin{table}[h]
		\begin{ruledtabular}
			\centering
			\caption{Characterization of the three parameter assignment strategies: the one with constant interface width (IWc), with variable interface width and all anisotropy in $\gamma$ (IWvG) and with variable interface width and all anisotropy in $\kappa$ (IWvK). In the latter, it is inconvenient to choose other value of $\gamma$ than $\gamma=1.5$. IW stands for interface width, other symbols have meaning as in the text.}
			\label{tab_models_comparison}
			\begin{tabular}{l|c|c|c|}\footnotesize
				& IWc & IWvG & IWvK \\ \hline
				fixed parameters & IW, $m$ & $\kappa, m$ & $\gamma, m$  \\
				varying parameters & $\gamma, \kappa$ & IW, $\gamma$ & IW, $\kappa$
			\end{tabular}
		\end{ruledtabular}
	\end{table}
	
	\subsubsection{Systems with pair-wise isotropic IE}
	In the system with $n$ phase fields, there are $n(n-1)/2$ possible pair-wise interfaces, each of which may have different (mean) interface energy $\sigma_{i,j}$. The indices $i,j$ denote interface between phase fields $\eta_i,\eta_j$. A set of parameters $m,\kappa_{i,j},\gamma_{i,j}, L_{i,j}$ (all scalars) is obtained by appropriate procedure  (depending on the strategy, see~S.I in Supplemental Material~\cite{Minar2021suppl}) so that the relations \eqref{eq_IE} and \eqref{eq_IW} are valid for each interface independently (equations~\eqref{eq_def_kappa}-\eqref{eq_def_L} hold for each interface ($i$-$j$)). Then, these are combined together to produce the model parameter fields $\kappa(\bm{r}),\gamma(\bm{r}),L(\bm{r})$:
	\begin{equation}
		\kappa(\bm{r}) = \frac{\sum_{i=1}^n\sum_{j>i}^n\kappa_{i,j}\eta_i^2\eta_j^2}{\sum_{i=1}^n\sum_{j>i}^n\eta_i^2\eta_j^2}
	\end{equation}
	\begin{equation}
		\gamma(\bm{r}) = \frac{\sum_{i=1}^n\sum_{j>i}^n\gamma_{i,j}\eta_i^2\eta_j^2}{\sum_{i=1}^n\sum_{j>i}^n\eta_i^2\eta_j^2} \,,
	\end{equation}
	\begin{equation}
		L(\bm{r})= \frac{\sum_{i=1}^n\sum_{j>i}^nL_{i,j}\eta_i^2\eta_j^2}{\sum_{i=1}^n\sum_{j>i}^n\eta_i^2\eta_j^2}
	\end{equation}
	
	which stand in place of $\kappa,\gamma$ in the functional in equation~\eqref{eq_def_totalE_ver1} and in place of $L$ in the governing equation~\eqref{eq_ACeq_governing}.
	
	Notice that from Table~\ref{tab_models_comparison} stems  that in IWvG is $\kappa(\bm{r})=\mathrm{const}$ by definition (i.e. all the $\kappa_{i,j}$s are equal) and similarly in IWvK $\gamma(\bm{r})=\mathrm{const}=1.5$ (i.e. all $\gamma_{i,j}$s are equal). 
	
	The free energy functional for IWc is then (those for IWvG and IWvK are equal, only with either $\kappa(\bm{r})$ or $\gamma(\bm{r})$ being constants, respectively):
	\begin{equation} \label{eq_def_totalE_ver2}
		F = \int_V \Bigg\{ m f_0(\vec{\eta}) + \frac{\kappa(\bm{r})}{2}\sum_{i=1}^n(\nabla \eta_i)^2 \Bigg\} \mathrm{d}V \,,
	\end{equation}
	\begin{equation}
		f_0(\vec{\eta}) = \sum_{i=1}^n\left(\frac{\eta_i^4}{4} - \frac{\eta_i^2}{2} \right) +\sum_{i=1}^n\sum_{i>j}\gamma_{i,j}\eta_i^2\eta_j^2 + \frac{1}{4} \,.
	\end{equation}
	
	Both parameter fields $\kappa(\bm{r}),\gamma(\bm{r})$ are functions of phase fields $\vec{\eta}$. This dependence should produce new terms in the governing equations (from $\partial f/\partial \eta_p$ in equation~\eqref{eq_ACeq_governing}). However, because the denominator of $\gamma(\bm{r})$ cancels out in the functional, the new terms only arise from $\partial \kappa/\partial \eta_p$.\\
	The governing equations then are 
	\begin{equation}
		\begin{split}
			\frac{\partial \eta_p}{\partial t} = -L(\bm{r})\left[ m\left( \eta_p^3-\eta_p +  2\eta_p\sum_{j\neq p}\gamma_{p,j}\eta_j^2 \right) \right. \\ \left. +\frac{1}{2}\frac{\partial \kappa}{\partial\eta_p}\sum_{i=1}^n(\nabla\eta_i)^2 - \kappa(\bm{r})\nabla^2\eta_p \right] 
		\end{split}
	\end{equation}
	
	The above procedure is fully variational, nevertheless inclusion of the term proportional to $\partial \kappa/\partial \eta_p$ enables the model to reduce the total energy of the system by introduction of so called third phase contributions (also ghost or spurious phases) at diffuse interfaces.~\cite{Moelans2008} That is a common problem in multi-phase field models~\cite{Toth2015}, where a third phase field attains non-zero value within an interface of two other phase fields. This mathematical artefact affects triple junction angles and in general is not physically justified. Several ways of elimination or suppression of ghost phases were described in~\cite{Toth2015} and the references therein.
	
	In this work, the ghost phases were eliminated by neglecting the term proportional to $\partial \kappa/\partial \eta_p$. However, because such model is not fully variational, the thermodynamic consistency can no longer be guaranteed in IWc and IWvK. This does not affect IWvG, because there is $\partial \kappa/\partial\eta_p=0$ anyway. That accounts for a clear advantage of the IWvG model, as no ghost phases appear even when fully variational.
	
	\subsubsection{Systems with inclination-dependent interface energy} \label{sec_model_incldepIE}
	\begin{table*}[]
		\centering
		\caption{Inclination dependence of the variable parameters in the respective models. The interface energy is $\sigma_{i,j}(\theta_{i,j})=\sigma_{i,j}^0h_{i,j}(\theta_{i,j})$. Symbols $\kappa_{i,j}^0, \gamma_{i,j}^0$ stand for scalar values of the parameters determined from $\sigma_{i,j}^0$ (see~S.I in~\cite{Minar2021suppl}). Expressions for $\gamma_{i,j}(\theta_{i,j})$ follow the so called \textit{weak anisotropy approximation}~\cite{Moelans2008}, i.e. they assume that the values of $\gamma_{i,j}(\theta_{i,j})$ do not diverge far from 1.5, so that the approximation $g^2[\gamma_{i,j}(\theta_{i,j})]\approx16[2\gamma_{i,j}(\theta_{i,j})-1]/9[2\gamma_{i,j}(\theta_{i,j}) +1]$ is applicable (see~\cite{Moelans2008} for details). Second row contains expressions used in equations~\ref{eq_dkppijdGp} and \ref{eq_dgmmijdGp}.}
		\label{tab_models_comp__par_incldep}
		\begin{ruledtabular}
			\begin{tabular}{p{2cm}|>{\centering\arraybackslash}p{4cm}>{\centering\arraybackslash}p{4.3cm}>{\centering\arraybackslash}p{3.5cm}}
				model     & IWc & IWvG & IWvK  \\ \hline
				variable parameter(s)     & $\begin{array}{l}
					\kappa_{i,j}(\theta_{i,j})= \kappa_{i,j}^0h_{i,j}(\theta_{i,j}) \\
					\gamma_{i,j}(\theta_{i,j})= -\frac{\frac{9}{4}g^2(\gamma_{i,j}^0)h_{i,j}(\theta_{i,j})+1}{\frac{9}{2}g^2(\gamma_{i,j}^0)h_{i,j}(\theta_{i,j})-2}
				\end{array}$
				& $\gamma_{i,j}(\theta_{i,j})= -\frac{\frac{9}{4}[g(\gamma_{i,j}^0)h_{i,j}(\theta_{i,j})]^2+1}{\frac{9}{2}[g(\gamma_{i,j}^0)h_{i,j}(\theta_{i,j})]^2-2}$
				&
				$\kappa_{i,j}(\theta_{i,j})= \kappa_{i,j}^0[h_{i,j}(\theta_{i,j})]^2$ \\ \hline
				$\partial \kappa_{i,j}/\partial h_{i,j}$ and $\partial \gamma_{i,j}/\partial h_{i
					,j}$ 
				& $\begin{array}{l}
					\partial \kappa_{i,j}/\partial h_{i,j} = \kappa_{i,j}^0 \\
					\partial \gamma_{i,j}/\partial h_{i,j} = \frac{9g^2( \gamma_{i,j}^0)}{\left[\frac{9}{2}g^2(\gamma_{i,j}^0)h_{i,j}(\theta_{i,j}) - 2\right]^2}
				\end{array}$ 
				& $\partial \gamma_{i,j}/\partial h_{i,j} = \frac{18g^2(\gamma_{i,j}^0)h_{i,j}(\theta_{i,j})}{\left\{\frac{9}{2}[g(\gamma_{i,j}^0)h_{i,j}(\theta_{i,j})]^2 - 2\right\}^2}$ 
				& $\partial \kappa_{i,j}/\partial h_{i,j} = 2 \kappa_{i,j}^0h_{i,j}(\theta_{i,j})$
			\end{tabular}
		\end{ruledtabular}
	\end{table*}
	The orientation of an interface in 2D system is given by interface normal, inclined under the angle $\theta$. Local value of interface energy may be a function of local interface inclination, i.e. $\sigma = \sigma(\theta)$. In Moelans' model~\cite{Moelans2008}, the normal at interface between $\eta_i,\eta_j$, denoted $\hat{\bm{n}}_{i,j}$, is defined as 
	\begin{equation}
		\hat{\bm{n}}_{i,j} = \frac{\nabla\eta_i-\nabla\eta_j}{|\nabla\eta_i-\nabla\eta_j|} = \left[\begin{array}{c}
			(\hat{n}_{i,j})_x   \\
			(\hat{n}_{i,j})_y
		\end{array} \right]
	\end{equation}
	and the definite inclination of that normal 
	\begin{equation}
		\theta_{i,j} = \mathrm{atan2}[(\hat{n}_{i,j})_y,(\hat{n}_{i,j})_x] \,,
	\end{equation}
	which is the standard 2-argument arctangent function.\\
	In 2D, the inclination-dependence of interface energy can be expressed as 
	\begin{equation}\label{eq_IE_incldep}
		\sigma_{i,j}(\theta_{i,j}) = \sigma_{i,j}^0h_{i,j}(\theta_{i,j})
	\end{equation}
	where $\sigma_{i,j}^0$ is a scalar and  $h_{i,j}(\theta_{i,j})$ is anisotropy function. The used anisotropy function was
	\begin{equation}
		h_{i,j}(\theta_{i,j}) = 1 + \delta\cos(n\theta_{i,j}) \,,
	\end{equation}
	
	with $\delta$ being strength of anisotropy and $n$ the order of symmetry. Some properties of this anisotropy function and the resulting Wulff shapes are given in~S.II of the Supplemental Material~\cite{Minar2021suppl}. 
	
	The inclination dependence of $\sigma_{i,j}$ implies that some of the model parameters $\gamma_{i,j},\kappa_{i,j}$ must be taken inclination-dependent too. Depending on the model used (IWc, IWvG or IWvK), the local validity of equation~\eqref{eq_IE} is achieved using different inclination dependence of the variable parameters (see Table~\ref{tab_models_comp__par_incldep} for details).
	
	Because the inclination-dependent $\kappa_{i,j},\gamma_{i,j}$ are functions of components of gradients $\nabla\eta_i,\nabla\eta_j$, the divergence term in the functional derivative (equation~\eqref{eq_ACeq_governing}) produces additional driving force terms. In the general case with multiple inclination-dependent interfaces, the divergence term equals
	\begin{equation}\label{eq_incldep_divDFterms}
		\begin{split}
			\nabla\cdot \frac{\partial f}{\partial(\nabla \eta_p)} &= 2m\eta_p\nabla\eta_p\cdot \left[\sum_{j\neq p}  \eta_j^2\frac{\partial \gamma_{p,j}}{\partial (\nabla\eta_p)}\right]  \\ 
			&\quad+ 2m\eta_p^2\sum_{j\neq p} \left[\eta_j\nabla\eta_j \cdot \frac{\partial \gamma_{p,j}}{\partial (\nabla\eta_p)}\right] \\
			&\quad+ m\eta_p^2\sum_{j\neq p}\eta_j^2\left[\nabla\cdot\frac{\partial \gamma_{p,j}}{\partial (\nabla\eta_p)}\right] \\
			&\quad + \frac{1}{2}\left[\nabla\cdot\frac{\partial \kappa}{\partial (\nabla\eta_p)}\right]\sum_{i=1}^n(\nabla \eta_i)^2 \\
			&\quad+ \frac{1}{2}\frac{\partial \kappa}{\partial (\nabla\eta_p)}\cdot\left[\nabla\sum_{i=1}^n(\nabla \eta_i)^2\right]  \\
			&\quad+ \nabla\kappa(\bm{r})\cdot\nabla\eta_p + \kappa(\bm{r})\nabla^2\eta_p \,.
		\end{split} 
	\end{equation}
	The vector field $\partial \kappa/\partial(\nabla\eta_p)$ is
	\begin{equation} \label{eq_dkppdGp}
		\frac{\partial \kappa}{\partial(\nabla\eta_p)} =  \frac{\sum\limits_{j\neq p}^n \left(\frac{\partial \kappa_{p,j}}{\partial (\nabla\eta_p)}\right)\eta_p^2\eta_j^2}{\sum\limits_{k=1}^n\sum\limits_{l>k}\eta_k^2\eta_l^2} \,
	\end{equation}
	where the sum in the numerator goes through all pair-wise interfaces of $\eta_p(\bm{r})$. The vector fields $\partial\kappa_{p,j}/\partial (\nabla\eta_p)$ are expressed
	\begin{equation} \label{eq_dkppijdGp}
		\frac{\partial \kappa_{p,j}}{\partial (\nabla\eta_p)} = \frac{1}{|\nabla\eta_i-\nabla\eta_j|}\frac{\partial \kappa_{p,j}}{\partial h_{p,j}}\frac{\partial h_{p,j}}{\partial \theta_{p,j}} \left[\begin{array}{c}
			-(\hat{n}_{i,j})_y   \\
			(\hat{n}_{i,j})_x
		\end{array} \right].
	\end{equation}
	Note, that the above vector field is nonzero only in IWc and IWvK models at the interfaces ($p$-$j$) with inclination-dependent IE. Likewise, the below vector field is nonzero only in IWc and IWvG
	\begin{equation} \label{eq_dgmmijdGp}
		\frac{\partial \gamma_{p,j}}{\partial (\nabla\eta_p)} = \frac{1}{|\nabla\eta_i-\nabla\eta_j|}\frac{\partial \gamma_{p,j}}{\partial h_{p,j}}\frac{\partial h_{p,j}}{\partial \theta_{p,j}} \left[\begin{array}{c}
			-(\hat{n}_{i,j})_y   \\
			(\hat{n}_{i,j})_x
		\end{array} \right].
	\end{equation}
	The multipliers $\partial \kappa_{p,j}/\partial h_{p,j}$ and $\partial \gamma_{p,j}/\partial h_{p,j}$ differ in individual models and are also provided in Table~\ref{tab_models_comp__par_incldep}. The term $\partial h_{p,j}/\partial \theta_{p,j}$ is defined by the inclination-dependence at the interface ($p$-$j$).
	
	The governing equation then is
	\begin{equation}
		\begin{split}
			\frac{\partial \eta_p}{\partial t} = -L(\bm{r})\left[ m\left( \eta_p^3-\eta_p +  2\eta_p\sum_{j\neq p}\gamma_{p,j}(\theta_{p,j})\eta_j^2 \right) \right. \\ 
			\left.  -\nabla\cdot \frac{\partial f}{\partial(\nabla \eta_p)} \right] \,.
		\end{split}
	\end{equation}
	
	Note, that the term proportional to $\partial \kappa/\partial\eta_p$ was neglected here. 
	
	In models with variable interface width (IWvG, IWvK), at the interfaces with inclination-dependent interface energy, the interface width is a function of the inclination, i.e. $l_{i,j}=l_{i,j}(\theta_{i,j})$. Because the kinetic coefficient $L_{i,j}$ is inversely proportional to the interface width $l_{i,j}$ (see equation~\eqref{eq_def_L}), the kinetic coefficient is inclination-dependent as well (even for \textit{constant} grain boundary mobility $\mu_{i,j}$). The inclination dependence of $L_{i,j}(\theta_{i,j})$ due to interface width variation is in the IWvG model
	\begin{equation}\label{eq_Lcorr_IWvG}
		L_{i,j}(\theta_{i,j}) = L_{i,j}h_{i,j}(\theta_{i,j})
	\end{equation}
	and in the IWvK model
	\begin{equation}\label{eq_Lcorr_IWvK}
		L_{i,j}(\theta_{i,j}) = L_{i,j}/h_{i,j}(\theta_{i,j}) \,,
	\end{equation}
	where $h_{i,j}(\theta_{i,j})$ is the anisotropy function in interface energy~\eqref{eq_IE_incldep}.
	
	The equations~\eqref{eq_Lcorr_IWvG},\eqref{eq_Lcorr_IWvK} were derived from an alternative expression for the kinetic coefficient $L_{i,j}$
	\begin{equation}
		L_{i,j} = \frac{\mu_{i,j}\sigma_{i,j}(\theta_{i,j})}{\kappa_{i,j}(\theta_{i,j})} \,,
	\end{equation}
	where the inclination dependencies of the right-hand side were expressed correspondingly to the model (see Table~\ref{tab_models_comp__par_incldep} for $\kappa_{i,j}(\theta_{i,j})$).
	
	Due to varying number of driving force terms in the three parameter assignment strategies, the governing equations are different in each and hence it is justified to call them different models.
	
	\subsubsection{Systems with inclination-dependent mobility}
	Let the interface ($i$-$j$) have isotropic interface energy and inclination-dependent grain boundary mobility with anisotropy function $h_{i,j}^\mu(\theta_{i,j})$, i.e. $\mu_{i,j}=\mu_{i,j}(\theta_{i,j})=\mu_{i,j}^0h_{i,j}^\mu(\theta_{i,j})$. From equation~\eqref{eq_def_L} we can see that the kinetic coefficient must have the same anisotropy, i.e. $L_{i,j}(\theta_{i,j})=L_{i,j}^0h_{i,j}^\mu(\theta_{i,j})$, where $L_{i,j}^0=\mu_{i,j}^0g(\gamma_{i,j})/l_{i,j}f_{0c}(\gamma_{i,j})$. 
	
	If the interface energy is inclination-dependent as well and a model with variable interface width is used (either IWvG or IWvK), the inclination dependence in $L_{i,j}(\theta_{i,j})$ due to the interface width variation must be included similarly like in \eqref{eq_Lcorr_IWvG} and \eqref{eq_Lcorr_IWvK}. The physical inclination-dependence is independent from the one due to interface width variation, implying the following expression for IWvG model
	\begin{equation}     
		L_{i,j}(\theta_{i,j}) = L_{i,j}^0h_{i,j}(\theta_{i,j})h_{i,j}^\mu(\theta_{i,j})
	\end{equation}
	and for the IWvK model analogically
	\begin{equation}
		L_{i,j}(\theta_{i,j}) = L_{i,j}^0\frac{h_{i,j}^\mu(\theta_{i,j})}{h_{i,j}(\theta_{i,j})} \,,
	\end{equation}
	where $h_{i,j}(\theta_{i,j})$ is the interface energy anisotropy function.
	
	\subsection{Interface profiles in different models} \label{sec_difference_in_profiles}
	The main difference in the model modifications is how the interface width varies as function of local interface energy. Obviously, in IWc the width is constant. In IWvK (with $\gamma=1.5$) the width of interface $i$-$j$ can be computed as
	\begin{equation}
		l_{i,j} = 6\frac{\sigma_{i,j}}{m} \,.
	\end{equation}
	Apparently, in IWvK model the interface width is proportional to the interface energy, i.e. the larger the interface energy, the larger the interface width.
	
	In IWvG the width can be expressed from~\eqref{eq_IW} and~\eqref{eq_IE} assuming $\frac{4}{3}\sqrt{f_{0,c}}(\gamma_{i,j})=g(\gamma_{i,j})$ (which holds for small values of $\gamma_{i,j}$). Then, it goes approximately
	\begin{equation}
		l_{i,j} \approx \frac{\kappa}{\sigma_{i,j}} \,,
	\end{equation}
	and apparently the larger interface energies are associated with lower interface widths in IWvG. 
	
	\section{Numerical implementation}
	All models were implemented in a single MATLAB function, where the governing equations were solved by centered finite differences of second order, explicit Euler time stepping and boundary conditions implemented using ghost nodes. The minimal code to run the simulations is available in the dataset~\cite{Minar2022dataset}. \\
	During the parameters assignment in models with variable interface width (IWvG, IWvK), there was assured control over the minimal interface width, i.e. that there would be no interface narrower than the user-specified one. That is to prevent unphysical behavior of the interface due to too small grid resolution. Different strategies had to be adopted in IWvK and IWvG, respectively. They are described in Supplemental Material~\cite{Minar2021suppl} together with other best practices in parameters determination for the respective models. The MATLAB functions which were used for parameters determination were also included in the dataset~\cite{Minar2022dataset}.
	
	The Supplemental Material~\cite{Minar2021suppl} further contains several practical details regarding the implementation such as time step determination as function of the anisotropy, the used finite-difference stencil and the driving force localization on the interface for solver stability.
	
	In the simulations with inclination-dependent interface energy, the vector field $\partial \kappa/\partial(\nabla\eta_p)$ was computed as in equations~\eqref{eq_dkppdGp} and \eqref{eq_dkppijdGp}, the fields $\partial \gamma_{p,j}/\partial(\nabla\eta_p)$ as in equation~\eqref{eq_dgmmijdGp} and their divergences (see equation~\eqref{eq_incldep_divDFterms}) were computed numerically (by centered differences), as well as the gradient $\nabla\kappa(\bm{r})$. All the above terms were computed in the IWc model, whereas in the models with variable interface width some of them could be omitted (as explained in section~\ref{sec_model_incldepIE}).
	
	When the anisotropy in inclination-dependent interface energy was strong (i.e. $\delta>1/(n^2-1)$ or $\Omega>1$), the anisotropy function had to be regularized as described in~\cite{Eggleston2001} in order to avoid ill-posedness of the governing equations for interfaces with missing inclination.
	
	\section{Methodology}
\begin{figure}[]
		\centering
		\includegraphics[page=8]{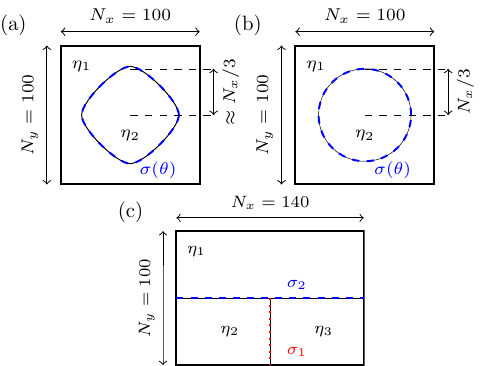}
		\caption{Initial conditions in the different numeric experiments with indicated interface energies. In a) Wulff shape shrinkage, b) kinetically compensated anisotropic curvature-driven circle shrinkage and c)  measurement of triple junction angle.  Grid dimensions correspond to the base run and 1000IW run (see text for details). All interfaces have equal mobilities.}
		\label{fig_IC_sketch}
	\end{figure}

	Three different simulation experiments were performed: a shrinking Wulff shape, kinetically compensated anisotropic curvature-driven circle shrinkage and triple junction angles. The initial-state geometries and grid dimensions are in Figure~\ref{fig_IC_sketch}. 
	\label{sec_Numeric}
	Except for the shrinking circles simulation, a parametric study was carried out in every experiment in order to validate the model behavior. Table~\ref{tab_overview_simulations} summarizes the variable parameters in every experiment.
	\begin{table*}[]
		\centering
		\caption{Overview of the simulations carried out in every simulation experiment. Note that these were carried out in every model modification (i.e. IWc, IWvG and IWvK) and simulation run (see Table~\ref{tab_IW_settings}). Number of simulations in every experiment is provided in the column Count. See text for more details.}
		\label{tab_overview_simulations}
		\begin{ruledtabular}
			\begin{tabular}{lc|l|l}
				Experiment          & Varied par.                   & Values                   & Count \\ \hline
				Wulff shape         & $\Omega$ (-)                  & 0.2, 0.4, 0.6, 0.8, 1.0, 2.3, 3.6, 4.9, 6.2, 7.5 & 10 \\ \hline
				Kin.comp.aniso. circle & $\Omega$ (-)                  & 0.1, 
				0.3, 0.5, 0.7 , 0.9       & 5 \\ \hline
				Triple junction     & $\sigma_1/\sigma_2$ (-)       & \begin{tabular}[c]{@{}l@{}}0.13, 0.2, 0.3, 0.4, 0.5, 0.6, 0.7, 0.8, 0.9, 1.0 \\ 1.1, 1.2, 1.3, 1.4, 1.5, 1.6, 1.7, 1.8, 1.9, 2.0\end{tabular}  & 20 
			\end{tabular}
		\end{ruledtabular}
	\end{table*}
	
	Each of the experiments was simulated using all the three model modifications (IWc, IWvG and IWvK) successively and the results were quantitatively compared. 
	
	In order to distinguish the behavior of the model from the artefacts of numeric discretization, the above described series of simulations was re-run for several numeric settings. Throughout the paper, these large series are called 'runs'. 
	
	Results of four runs are presented in this paper, the basic settings of which are summarized in Table~\ref{tab_IW_settings}. The difference between the runs were: a) the minimal set interface width $l_{min}$ and b) the number of points in the interface. The runs are denoted base run, IW/2, 14IWpts and 1000IW. The base run had rather coarse grid and 7 points in the interface, which should in general be reliable yet not very computationally heavy settings. The minimal interface width was $l_{min}=1$~nm. In the run IW/2 the interface width was halved ($l_{min}=0.5$~nm) and the number of points in the interface was 7 again. In the 14IWpts run, the interface width was as in the base run but the points in the interface were doubled (to be 14). Physical dimensions of the simulated domain were equal in the described runs but the grid spacing was halved in IW/2 and 14IWpts runs. In the last 1000IW run, the grid was equal as in the base run, but the physical dimensions were scaled by a factor of 1000, meaning the minimal interface width $l_{min}=1\,\mathrm{\mu m}$. This is the practical settings for the actual grain growth simulations. The latter run validated that the meso-scale behavior of the model is equal to that one at the nanometer scale.
	\squeezetable
	\begin{table}[]
		\centering
		\caption{Numeric settings of different simulation runs (for each simulation experiment and model). $l_{min}$ is the minimal interface width. $N_x$ and $N_y$ are grid dimensions in the base run as shown in Figure~\ref{fig_IC_sketch}}
		\label{tab_IW_settings}
		\begin{ruledtabular}
			\begin{tabular}{c|c|c|c|c}
				& base run & IW/2 & 14IWpts & 1000IW \\ \hline
				grid dimensions & $N_x$ x $N_y$ & $2N_x$ x $2N_y$ & $2N_x$ x $2N_y$ & $N_x$ x $N_y$ \\
				$l_{min}$ (nm) & 1 & 0.5 & 1 & 1000 \\
				points in $l_{min}$ & 7 & 7 & 14 & 7
			\end{tabular}
		\end{ruledtabular}
	\end{table}
	
	In the simulation experiments, position of the interface in the respective geometry was compared to the expected shape. Depending on the parameters assignment strategy, the phase field profiles may not be symmetric about the point 0.5 in Moelans' model. For this reason, it was considered that the position of the interface $i$-$j$ was in points $\bm{r}=(x,y)$ of the domain, where $\eta_i(\bm{r})=\eta_j(\bm{r})$, i.e. where the two profiles crossed. This way the contours were always well defined and could be quantitatively compared to appropriate analytical models even for Wulff shape simulations, where the profile shape varies along the interface.
	
	The below subsections describe in detail the analytic solutions of the problems and results processing in the individual simulation experiments. 
	
	\subsection{Used anisotropy function and its Wulff shape}
	The anisotropy function $h(\theta)=1+\delta\cos(n\theta)$ was used, together with the normalized strength of anisotropy $\Omega=\delta(n^2-1)$. $\Omega$ easily distinguishes weak from strong anisotropy, because for $0<\Omega<1$ the Wulff shape is smooth, whereas for $1\leq\Omega<n^2-1$ it has corners. The Supplemental material~\cite{Minar2021suppl} provides more details about this anisotropy function. Fourfold symmetry was assumed (i.e. $n=4$).
	
	The Wulff shape $\bm{w}$ in 2D is a planar curve, which can be parametrized by the interface normal angle $\theta$, i.e.  $\bm{w}(\theta)=[w_x(\theta),w_y(\theta)]^\mathrm{T}$, giving \cite{Burton1951,Kobayashi2001,Eggleston2001}
	
	\begin{align} 
		w_x(\theta) &= R_W[h(\theta)\cos(\theta) - h'(\theta)\sin(\theta)] \label{eq_wulff_parametrically_x}\\
		w_y(\theta) &= R_W[h(\theta)\sin(\theta) + h'(\theta)\cos(\theta)] \label{eq_wulff_parametrically_y}\,,
	\end{align}
	where $R_W>0$ is the radius of the Wulff shape and $h(\theta),h'(\theta)$ the anisotropy function and its derivative, respectively.  
	
	Ill-posedness of the governing equations for forbidden orientations on Wulff shapes for strong anisotropies ($\Omega>1$) was treated by regularization of the anisotropy function as in~\cite{Eggleston2001}.
	
	With $h(\theta)=1+\delta\cos(n\theta)$, the minimal distance $R_{min}$ between the Wulff shape center and the contour can be related to $R_W$ as
	\begin{equation} \label{eq_appdx_wulff_minradius}
		R_{min} = R_W(1-\delta) \,,
	\end{equation}
	which holds for arbitrarily strong anisotropy because the minimal-radius point normal is always inclined under a non-missing angle. This formula was used to find the radius $R_W$ of the phase field contour. Then, the phase field contour was scaled to unit radius and compared to the analytic Wulff shape (by means of Hausdorff distance, see the next subsection). 
	
	For validation of the kinetics of shrinkage, an analytic expression for Wulff shape shrinkage rate was derived in~S.III of Supplemental Material~\cite{Minar2021suppl}. Measurement of area/volume occupied by a grain/phase is trivial in phase field method, hence the rate of its change (i.e. the shrinkage rate) can be easily used for validation or benchmark. The derivation in~\cite{Minar2021suppl} delivers the expression
	\begin{equation}\label{eq_wulff_shrrate}
		\frac{\mathrm{d}A_W}{\mathrm{d}t} = -2\pi\mu\sigma_0  \frac{C_W(\Omega,n)}{1-\delta} \,
	\end{equation}
	where $C_W(\Omega,n)=A_W/A_{circle}$ is an anisotropic factor relating the area of a Wulff shape and a circle of equal radius. For fourfold symmetry it was numerically computed and fitted by polynomial $C_W(\Omega,4)= \sum_{i=0}^4a_i\Omega^{N-i}$ with $a_0=-0.00032 ,a_1=0.00639 ,a_2=-0.04219 , a_3=0.00034 , a_4=1.00000$. As can be seen, the analytic shrinkage rate is a constant, which is consistent with~\cite{Taylor1998}.
	
	With isotropic interface energy ($\delta=\Omega=0$) the steady state shape is a circle and the anisotropic factor is $C_W/(1-\delta)=1$, hence the isotropic curvature-driven shrinkage rate of a circle is (as also e.g. in~\cite{Moelans2009})
	\begin{equation} \label{eq_iso_shrrate}
		\frac{\mathrm{d}A}{\mathrm{d}t} = -2\pi\mu\sigma_0 \,.
	\end{equation}
	
	\subsection{Quantifying the match in shape}
	The \textit{Hausdorff distance} was used for quantification of the match in shape. Let the 2D curves $\bm{w},\bm{w}_{PF}$ be the analytic shape and the phase-field contour, respectively. The Hausdorff distance $d_H(\bm{w},\bm{w}_{PF})=\lambda$ between them implies, that $\lambda$ is the smallest number such that $ \bm{w} $ is completely contained in $\lambda$-neighborhood of $\bm{w}_{PF}$ and vice-versa~\cite{Alt2004}). Formally, it is defined between sets $P$ and $Q$ as 
	\begin{equation}
		d_H(Q,P)=\mathrm{max}(\tilde{d}_H(P,Q),\tilde{d}_H(Q,P)) \,,
	\end{equation}
	where 
	\begin{equation}
		\tilde{d}_H(P,Q)=\underset{x\in P}{\mathrm{max}}( \underset{y\in Q}{\mathrm{min}}||x-y|| )
	\end{equation}
	is \textit{directed Hausdorff distance}. It is always $d_H(\cdot,\cdot)\geq 0$, and the closer to zero, the more alike the compared sets are. It has been extensively used for image matching and pattern recognition~\cite{LiZhu2014}. 
	
	For comparability in the two validation experiments with inclination-dependent interface energy, it is essential that the two curves $\bm{w},\bm{w}_{PF}$ are co-centric and scaled to unit radius.
	
	Note tat the data set~\cite{Minar2022dataset} includes also MATLAB functions for the contour shape matching.
	
	\subsection{Quantifying match in shrinkage rate}
	The shrinkage rate was obtained as mean value of shrinkage rates in simulation time interval where the area of the shape was in between 0.95-0.6 fraction of the initial area. This choice should prevent the diffuse interface from being too large compared to the shape itself, in which case it would affect the kinetics. Additionally, this approach turned out to be rather insensitive to the particular numeric settings, which is convenient for validations. 
	
	The results are presented as relative error $\delta x$, defined in the following convention
	\begin{equation}
		\delta x = 100 \frac{x_0-x}{x}~\%\,,
	\end{equation}
	where $x_0$ is the measured value and $x$ is the expected one. In this convention, the shrinkage was \textit{slower} than expected when $\delta x<0$, and \textit{faster} when $\delta x > 0$.
	
	\subsection{Wulff shape}
	
	Shrinking Wulff shapes with different strengths of anisotropy were simulated. The match to the analytic shape was measured in Hausdorff distance. The shrinkage rate was expressed analytically and used for validation as well.
	
	The Neumann boundary conditions were applied to all boundaries in a system with initial condition like in Figure~\ref{fig_IC_sketch}a. The initial condition in every simulation was the analytic Wulff shape of the corresponding strength of anisotropy $\Omega$ as discretized by the grid. The radius was taken such that the initial shape occupied the area fraction in the domain of at least 0.25. Because the initial Wulff shape already minimized the interface energy, the shrinkage with constant rate as in~\eqref{eq_wulff_shrrate} was expected and any change in the shape was a departure from the analytic solution. 
	
	\subsection{Kinetically compensated anisotropic circle shrinkage}
	This simulation experiment validates the inclination-dependence of the kinetic coefficient in combination with inclination-dependent interface energy. 
	Specifically, the anisotropy of kinetic coefficient was chosen such that it compensated the anisotropic driving force so that the resulting interface motion was isotropic. 
	
	Again, the match to the steady-state shape (a circle) was quantified by Hausdorff distance and the mean shrinkage rate was measured when the shape area was a fraction 0.95-0.6 relative to the initial condition. Parametric study in $\Omega$ were carried out to validate the model.
	
	The initial condition for the simulation experiment was as in Figure~\ref{fig_IC_sketch}b, i.e. a two-phase-field system of a circular grain in a matrix.  
	
	Normal velocity $v_n$ of a curvature-driven interface with inclination-dependent interface energy is 
	\begin{equation}\label{eq_normal_velocity_aniso}
		v_n(\theta) = \frac{\mu}{\varrho}  \sigma_0[h(\theta)+h''(\theta)]\,,
	\end{equation}
	where $\mu$ is interface mobility, $\varrho$ is local radius of curvature and $\sigma_0[h(\theta)+h''(\theta)]$ is the interface stiffness. With $h(\theta)=1+\delta\cos(n\theta)$ the inclination-dependent factor in \eqref{eq_normal_velocity_aniso} is $[h(\theta)+h''(\theta)] = 1 - \delta(n^2-1)\cos(n\theta) = 1-\Omega\cos(n\theta)$.
	
	When the interface mobility is set anisotropic as
	\begin{equation} \label{eq_kincoeff_companiso}
		\mu(\theta) = \frac{\mu_0}{1-\Omega\cos(n\theta)} \,,
	\end{equation}
	the resulting interface normal velocity $v_n$ does not depend on interface inclination $\theta$ anymore, i.e. it is isotropic. The shrinkage rate is then~\eqref{eq_iso_shrrate}.
	
	The below presented simulations were all carried out with $\Omega<1$, because the kinetic coefficient as in equation~\eqref{eq_kincoeff_companiso} is then positive for all interface inclinations. 
	
	The ratio of maximal to minimal interface velocity due to the anisotropic interface energy is $(1+\Omega)/(1-\Omega)$, which indicates rather strong kinetic anisotropy when $\Omega$ is close to 1. E.g. with $\Omega=0.9$ the ratio of maximal to minimal interface velocity is 19 (assuming constant $\varrho$ for all inclinations, which holds for a circle). Note that the corresponding ratio of maximal to minimal interface energy is only 1.0664 (with four-fold symmetry).
	
	\subsection{Triple junction angles}
	
	In this experiment triple junction angles are measured in systems with different combinations of pair-wise isotropic interface energies. This way it is validated how well the triple junction force balance is reproduced by the model.
	
	The initial geometry was like in Figure~\ref{fig_IC_sketch}b with periodic left and right boundaries, and Neumann boundary conditions on the top and bottom ones. The individual interfaces are isotropic but have different interface energies. The initially straight interface segments (1-2) and (1-3) with grain boundary energy $\sigma_2$ turn into circular arcs, which then move towards the center of curvature, i.e. downwards. The two grains $\eta_2$ and $\eta_3$ will shrink and in the steady state the angle $\alpha$ between the arcs (see Figure~\ref{fig_trijunang_results}a) in the triple junction is described by Young's law~\cite{Porter2009} (section 3.3.3):
	\begin{equation}
		\alpha = 2\mathrm{acos}(\sigma_{1}/2\sigma_{2}) \,.
	\end{equation}
	
	The ratio $\sigma_1/\sigma_2$ was varied in the parametric study to validate the model (see Table~\ref{tab_overview_simulations}). It was always $\sigma_1=0.3\,\mathrm{J/m^2}$ and $\sigma_2$ was computed from the ratio.
	
	The phase field contours of interfaces (1-2) and (1-3) were analyzed by two methods in order to determine the triple junction angle $\alpha$. First, the points on both the arcs nearest to the triple junction were fitted by a straight line (indicated by red segments in Figure~\ref{fig_trijunang_results}a) and second, the remaining arc points were fitted by a circular arc (see green segments in Figure~\ref{fig_trijunang_results}a). Simple geometric construction allows to determine the angle $\alpha$ from the fitted parameters in the latter case~\cite{Moelans2009} as
	\begin{equation}
		\alpha = 2\mathrm{acos}(x/R)\,,
	\end{equation}
	where $R$ is the fitted circle arc radius and $x$ is the horizontal distance from the triple junction (see the scheme in Figure~\ref{fig_trijunang_results}a). Width of the interval in which the arcs were fitted by straight lines (width of pink rectangles in Figure~\ref{fig_trijunang_results}a) was set to width of the interface (2-3) (i.e. half of the width on each side).\\
	Accuracy of the lines fitting is affected especially by the width of the above interval and that of the circle arc fitting is mostly affected by the simulated arc shape.
	
	\section{Results}
	\label{sec_Results}
	
	\subsection{Wulff shapes}
	\begin{figure}
		\centering
		\includegraphics[page=4]{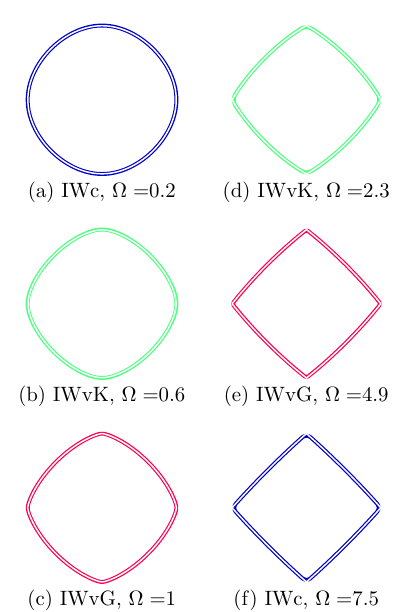}
		\caption{Demonstration of the simulated Wulff shapes for strengths of anisotropy $\Omega$ with the different models (base run). The white line is the analytic Wulff shape and the colored ones are the extracted phase field contours.}
		\label{fig_wulff_demo_shapes}
	\end{figure}
	
	\begin{figure}[]
		\centering
		\includegraphics[page=5]{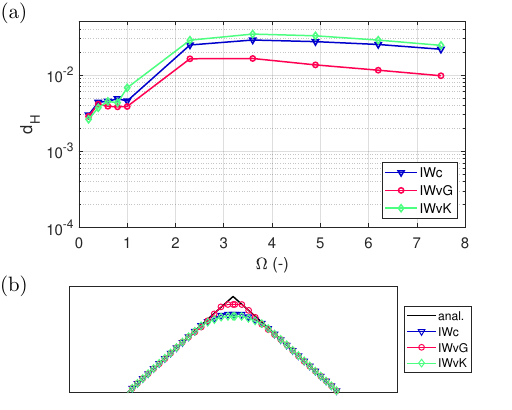}
		\caption{In (a) the match to Wulff shape for the model modifications in the base run as function of normalized strength of anisotropy $\Omega$. In (b) a detail of the phase field contours near  the Wulff shape corner for simulation with $\Omega=7.5$.}
		\label{fig_wulff_match}
	\end{figure}
	
	\begin{figure}
		\centering
		\includegraphics[page=6]{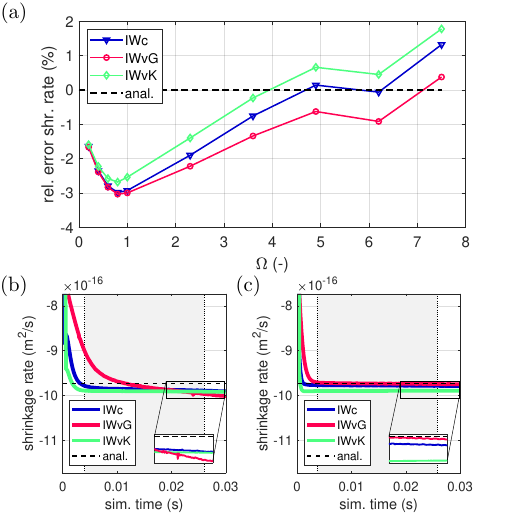}
		\caption{Wulff shape shrinkage rate results. In (a) the mean shrinkage rate in the base run as function of normalized strength of anisotropy $\Omega$, in (b) and (c) there is time evolution of shrinkage rate for $\Omega=7.5$ in the base and IW/2 runs, respectively. The shaded areas in b) and c) indicate the time interval from which the mean shrinkage rate was computed.}
		\label{fig_wulff_shrrate}
	\end{figure}
	
	The Figures~\ref{fig_wulff_demo_shapes}a-\ref{fig_wulff_demo_shapes}f visually compare the Wulff shapes obtained from simulation by the different models (in base run) to the analytic ones. As can be seen, the overall match is very good in all the three models, although a rather round contour near the corners in strong-anisotorpy Wulff shapes are observed (this is more apparent in Figure~\ref{fig_wulff_match}b, which shows detail of the contours near a corner for $\Omega=7.5$). That was expected, as no special finite difference scheme was used near corners (in~\cite{Eggleston2001} a one-sided finite difference scheme was proposed to avoid corners rounding). 
	
	Figure~\ref{fig_wulff_match}a, shows the match to Wulff shape as function of normalized strength of anisotropy $\Omega$ in the base run. As can be seen, the IWvG model slightly outperformed the other two in strong anisotropies because it was able to resolve the corners the best (see Figure~\ref{fig_wulff_match}b). However, when the interface width was halved in the IW/2 run, all the models performed nearly equally well because with smaller interface width the rounding near the corners was reduced. Interestingly, the IWvG model performed comparably in the IW/2 and base runs, which is in contrast to IWc and IWvK, which improved markedly with narrower interface width. \\
	The best results in match to Wulff shape were obtained in the 14IWpts run with IWvG model. IWc and IWvK models performed comparably in the 14IWpts and IW/2 runs.
	
	The mean shrinkage rates of the Wulff shapes as function of strength of anisotropy are in Figure~\ref{fig_wulff_shrrate}a for the base run. At first sight, all the models perform comparably well, following the analytical prediction within an absolute error of 3~\%. The Figures~\ref{fig_wulff_shrrate}b and \ref{fig_wulff_shrrate}c show the time evolution of shrinkage rate with $\Omega=7.5$ in the base and IW/2 runs, respectively. It can be seen that despite the mean shrinkage rates being near the prediction, in the base run, the shrinkage rate of IWvG model did not converge to the analytic prediction (see the inset of Figure~\ref{fig_wulff_shrrate}b, where the IWvG curve clearly declines). In the IW/2 run the IWvG model did converge close to the anaytic shrinkage rate (see also the inset of Figure~\ref{fig_wulff_shrrate}c) and all the mean values are a little closer than in the base run. Note that in the IWvG model the lowest-energy interface has the widest interface width and that the cornered Wulff shape contains only interface orientations with lower energy. For this reason it required narrower interface width to reach constant shrinkage rate.
	
	The relative error in mean shrinkage rate obtained with this methodology was nearly the same in the base and IW/2 runs though (both within $\pm3\,\%$). Apparently, the difference is that with narrower interface the attained shrinkage rate is more steady. Additional run with even finer grid was carried out and no improvement in the mean values of shrinkage rates was observed. It can thus be concluded that the methodology is robust enough to assess the shrinkage rate even in the domain 100x100 (i.e. base run).
	
	In addition to the four simulation runs discussed so far, the Wulff shapes simulations were re-run also with 4 and 5 points in the interface. The match to Wulff shape was worse than with 7 or 14 points, but the shapes were resolved qualitatively well regardless. The shrinkage rates were smaller than expected though, due to grid pinning. Seven points in the interface were thus confirmed as a reasonable value for the validations and practical simulations. 
	
	No significant effect of interface width scaling in the run 1000IW was found.
	
	\subsection{Kinetically compensated anisotropic circle shrinkage}
	
	\begin{figure}[]
		\centering
		\includegraphics[page=7]{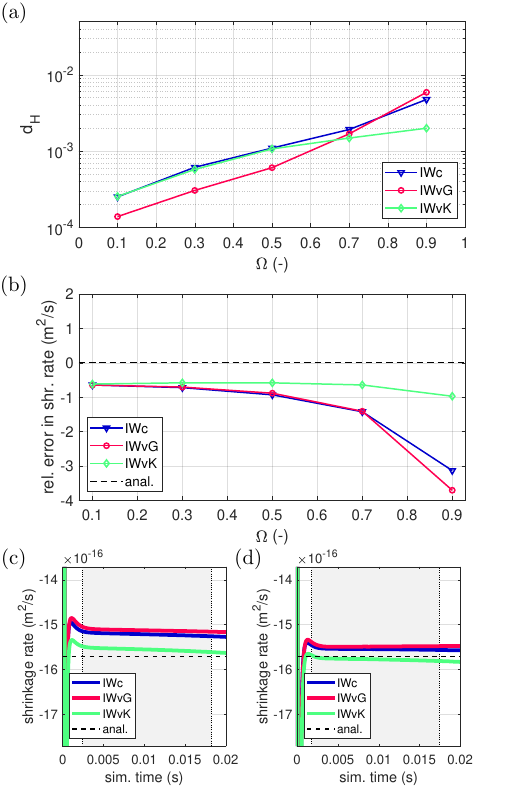}
		\caption{Results for kinetically compensated anisotropic shrinkage. In (a) the match to circle for the base run, in (b) the mean shrinkage rate for 14IWpts run, in (c) and (d) the shrinkage rate time evolution in simulation with $\Omega=0.9$ in the base run and 14IWpts runs, respectively.}
		\label{fig_companiso}
	\end{figure}
	
	The quantified match to the circle and the mean shrinkage rate as functions of strength of anisotropy $\Omega$ are in Figures~\ref{fig_companiso}a and \ref{fig_companiso}b, respectively (results of the base run showed). All models and runs retained the initial circle well or up to excellent geometrical match. Nevertheless, the IWvG model gave the best results, except for the strongest anisotropy, where the IWvK model was better. Only minor improvement in the match was achieved in the IW/2 run when compared to the base run. As with Wulff shapes simulations, the best match was obtained in the 14IWpts run, which also exhibited the best mean shrinkage rates (for all the three models). \\
	The relative error in shrinkage rates in Figure~\ref{fig_companiso}b shows slightly decreasing trend with $\Omega$ for IWc and IWvG models (i.e. slowing down). The values of IWvK model were not affected and were constant. Apparently, the symmetric profiles of IWvK model provide an advantage for preserving the expected kinetics in simulations with strong kinetic anisotropy.\\
	Time evolution of shrinkage rate for the strongest considered anisotropy (i.e. $\Omega=0.9$) in Figures~\ref{fig_companiso}c and \ref{fig_companiso}d (base and 14IWpts runs, respectively), shows that in both runs the lines slightly diverge (this being applicable to all $\Omega$s). The convergence was better in IW/2 run, but the mean shrinkage rates were worse than in the 14IWpts run.
	
	Apparently, optimal results would be obtained here with more points than 7 in the interface and with smaller interface-width-to-circle ratio than in the base run and 14IWpts runs. However, as noted earlier, the kinetic anisotropy is very strong in the case of $\Omega=0.9$. For weaker anisotorpy the discussed effects are less pronounced and there is little difference in the shrinkage rates among the models.
	
	There was no significant difference between the results of base and 1000IW runs.

	\begin{figure*}[]
		\centering
		\includegraphics[page=3]{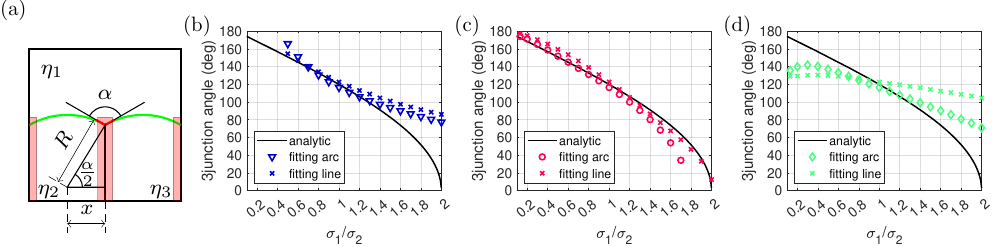}
		\caption{Triple junction angles. In a) the two methods for angles determination are illustrated (points fitted by straight lines in red and those fitted by circular arc in green). Subfigures b)-d) show the simulation results for different models, these being: in b) IWc, in c) IWvG and in d) IWvK. The hollow symbols correspond to the the angles determined by arc fitting and crosses to the lines fitting.}
		\label{fig_trijunang_results}
	\end{figure*}

	\subsection{Equilibrium triple junction angles}
	Figures~\ref{fig_trijunang_results}b-\ref{fig_trijunang_results}d show the simulation results from the triple junction simulations as function of $\sigma_1/\sigma_2$. Neither of IWc or IWvK models show good agreement to Young's law when deviating farther from an isotropic system (which has ratio $\sigma_1/\sigma_2=1$). The model IWvG, on the other hand, always shows very good agreement in at least one of the fitting methods along the whole range of probed ratios of interface energies. With $\sigma_1/\sigma_2$ closer to 2 in IWvG modification, the grains shape was slightly elongated in the vertical direction, resulting in too small radius of the fitted arcs to cross in the triple junction. The angles could not be determined this way then and the linear fit is more reliable. \\
	For the ratios approximately $\sigma_1/\sigma_2\leq0.45$ the IWc model behaves non-physically. The triple junction was observed to move in the opposite than expected direction (i.e. upwards, as if the triple junction angle was larger then 180\textdegree). The overall shape of contours was not as in Figure~\ref{fig_trijunang_results}a because the green arcs curved in the other way. In IWvK this behavior was not observed, but the Young's law is not followed. Qualitative explanation is that the IWc and IWvK models are not fully variational. \\
	No significant change was observed in the results of the triple junction angles in runs IW/2, 14IWpts or 1000IW when compared to the base run. Quality of the results is thus not improved when more points in the interface than 7 are used. Also it implies, that the model behavior (for all parameters assignment strategies) is not affected by reducing the interface-width-to-feature ratio or the interface width scaling. The latter confirms that the model is quantitative.
	
	\section{Conclusions}
	
	This paper presented a quantitative methodology for assessment of the anisotropic curvature driving force in phase field method. It was demonstrated in comparison of three different modifications of a multi-phase field model. The match to the expected shape and the shrinkage rate were quantified in two different benchmark problems. The methodology was sensitive enough to capture differences between the model modifications and is suitable for validation and benchmarking of different models and numerical solvers.
	
	The overall performance of the three model modifications in the benchmarks was comparable. Both match to the steady-state shapes and the shrinkage rates followed the expected results as in the anisotropic mean curvature flow. However, a significant difference was noted in a supplementary benchmark simulation where triple junction angles were measured. It was observed that only the IWvG model modification (with only the parameter $\gamma$ anisotropic) reproduced the triple junction angles in the full interval 0.13-2 of $\sigma_1/\sigma_2$, whereas IWc and IWvK modifications failed for ratios farther from 1. It is noted that these two modifications were not fully variational in order to avoid ghost phases, whereas IWvG is fully variational and yet the ghost phases do not appear.
	
	Even though the triple junction benchmark did not involve interfaces with inclination-dependent interface energies, one conclusion can be made about that case regardless. As the IWc and IWvK were shown unreliable in the simpler pair-wise isotropic case, there is no reason why they should be reliable in the more complicated one. Further development and validations should thus focus on the IWvG model modification.

	No results were significantly affected by the interface width scaling. Also, it was confirmed that 7 points in the interface were sufficient for retaining the expected kinetics in most cases, unless the inclination dependence of the kinetic coefficient was very strong.
	
	The results in this paper are reproducible with codes provided in the data set~\cite{Minar2022dataset}.
	
	\section{Acknowledgements}
	The authors acknowledge the support of European Research Council (ERC) under the European Union’s Horizon 2020 research and innovation program (INTERDIFFUSION, Grant Agreement No. 714754).


	
	%

	
\end{document}


\renewcommand{\thefigure}{S.\arabic{figure}}
	\renewcommand{\thetable}{S.\arabic{table}}
	\renewcommand{\theequation}{S.\arabic{equation}}
	\renewcommand{\thesection}{S.\Roman{section}}
	
	\preprint{APS/123-QED}
	
	\title{Supplemental Material for\\ Benchmark problems for anisotropic curvature-driven shrinkage: a multi-phase field model study}
	
	\author{Martin Minar}
	\email{martin.minar1132@gmail.com}
	\author{Nele Moelans}%
	
	\affiliation{KU Leuven, Deptartment of Materials Engineering}
	
	\date{\today}
	\maketitle
	
	\section{Model parameters determination - best practices}
	\label{sec_suppl_param_det}
	The best practices in parameters determination described in this section were implemented in MATLAB functions, which were published in dataset~\cite{Minar2022dataset}, distributed under GPLv3 license.
	
	\subsection{Polynomial expressions for determination of $\gamma$} \label{sec_param_det_polynomials}
	Interface energy and width are non-analytic functions of the parameter $\gamma$  (equations~(6)~and~(7) in the main text). In models IWc and IWvG with non-uniform interface energy, $\gamma$ varies in space and depends on physical input (local interface energy $\sigma$) and chosen interface width. Assuming that $\kappa,m$ and $\sigma$ are known, $g(\gamma)$ is computed using equation~(6) and must be inverted in order to obtain $\gamma$. For numeric convenience, functions $g^2\rightarrow 1/\gamma$ and $ 1/\gamma \rightarrow f_{0,c}$ were fitted by polynomials using the tabulated values from~\cite{Ravash2017}. The same approach was taken in~\cite{Moelans2008}. Note that in one of the following subsections an alternative algorithm for determination of parameters in IWc is proposed. \\
	The values of $f_{0,c}(\gamma),g(\gamma)$ were tabulated and published in~\cite{Ravash2017} in the range $0.52\leq\gamma\leq 40$. However, the polynomials used in both~\cite{Moelans2008,Ravash2017} did not describe the functions along the whole tabulated range. For inconvenient input, that resulted in a loss of control over the interface properties (even negative values of $\gamma$ were obtained). \\
	To improve the parameters determination, the functions $g^2\rightarrow 1/\gamma$ and $ 1/\gamma \rightarrow f_{0,c}$ were fitted again so that all $\gamma$ values from the interval $0.52\leq\gamma\leq 40$ were reliably available. The new polynomials show good agreement along the given interval and their coefficients are listed in Table~\ref{tab_param_polynomial_coeff_myfits}. 
	
	\begin{table*}[h]
		\centering
		\caption{Coefficients of the new polynomials used in the parameters determination algorithm (IWc and IWvG models). Based on fitting of the numeric values given in \cite{Ravash2017}. The polynomial of order $N$ $p_N(x)$ is defined here together with its coefficients as $p_N(x)=\sum_{i=0}^N a_i x^{N-i}$. See text for details of use in the respective models.}
		\label{tab_param_polynomial_coeff_myfits}
		\begin{ruledtabular}
			\begin{tabular}{c|c|ccccccc}
				& order & $a_0$ & $a_1$ & $a_2$ & $a_3$ & $a_4$ & $a_5$ & $a_6$  \\ \hline
				$g^2 \rightarrow \frac{1}{\gamma}$ & 6 & -5.73008 &  18.8615 & -23.0557 & 7.47952 & 8.33568 & -8.01224 & 2.00013 \\
				$\frac{1}{\gamma} \rightarrow \sqrt{f_{0,c}}$&   5 & -0.072966 & 0.35784 & -0.68325 & 0.63578  &  -0.48566   & 0.53703 &    \\
				$g\sqrt{f_{0,c}}\rightarrow \frac{1}{\gamma}$ & 6 & 103.397 & -165.393 &  105.3469 & -44.55661 &  24.7348 & -11.25718 & 1.999642
			\end{tabular}
		\end{ruledtabular}
	\end{table*}
	
	\subsection{IWvG}
	This parameters determination algorithm~\cite{Ravash2017} requires input of initializing interface energy $\sigma_{init}$, and width $l_{init}$ which are assigned to interface with value $\gamma=1.5$. There, the analytic relations between the model parameters are valid and the (constant) parameters $\kappa,m$ can be computed:
	\begin{equation}\label{eq_anal_kappa_m}
		\kappa = \frac{3}{4}\sigma_{init}l_{init} \quad ,\quad m = 6\frac{\sigma_{init}}{l_{init}} \,.
	\end{equation}
	These are then used in equation~(6) with individual interface energies $\sigma_{i,j}$ to determine $g^2(\gamma_{i,j})$ and further $\gamma_{i,j}$ (using the first polynomial from Table~\ref{tab_param_polynomial_coeff_myfits}).\\
	However, because the said polynomial was only fitted in the interval $0.52\leq\gamma\leq 40$ (values tabulated in~\cite{Ravash2017}), it is important to choose the value of $\sigma_{init}$ such, that none of the interface energies $\sigma_{i,j}$ should be described by $\gamma_{i,j}$ out of the interval. From equation~(6) with substituted $\sqrt{\kappa m}$ from \ref{eq_anal_kappa_m} we show that the values do not leave the above interval when $\sigma_{init}$ complies with the following two inequalities
	\begin{equation}
		\frac{\sigma_{max}}{g(40)}\sqrt{\frac{2}{9}}\leq \sigma_{init}\leq \frac{\sigma_{min}}{g(0.52)}\sqrt{\frac{2}{9}} \,,
	\end{equation}
	where $\sigma_{min},\sigma_{max}$ are minimal and maximal interface energy present in the system. Additionally, in order to fulfill these conditions, it must be satisfied that
	\begin{equation}
		\frac{\sigma_{min}}{\sigma_{max}} \geq \frac{g(0.52)}{g(40)} = \frac{0.1}{0.76} = 0.129 \,,
	\end{equation}
	which sets the maximal anisotropy in interface energy for which all $\gamma_{i,j}$s may be determined in IWvG based on the values tabulated in~\cite{Ravash2017}.\\
	Control over the width of the narrowest interface (with user-defined width $l_{min}$) is obtained when the parameters are determined in two steps as follows:
	\begin{enumerate}
		\item values of $\kappa,m$ are calculated from equations~\eqref{eq_anal_kappa_m} using appropriate $\sigma_{init}$ together with $l_{min}$
		\item values of $\kappa,m$ are recalculated from equations~\eqref{eq_anal_kappa_m} with the same $\sigma_{init}$ and interface width $l_{calib}=l_{min}\sqrt{8f_{0,c}(\gamma_{min})}$,  where $\gamma_{min}$ corresponds to the lowest value of all the $\gamma_{i,j}$ obtained in the previous step, i.e. to the interface with the largest interface energy $\sigma_{max}$. 
	\end{enumerate}
	After the second step, the interface with maximal interface energy will have the width $l_{min}$ and all the other interfaces will have it equal or larger.
	
	The other interface widths can be expressed
	\begin{equation}
		l_{i,j} = \frac{\sigma_{i,j}}{m}\frac{1}{g(\gamma_{i,j})\sqrt{f_{0,c}}(\gamma_{i,j})}
	\end{equation}
	
	\subsection{IWc}
	The iterative parameters determining procedure assuring the constant interface width is described in~\cite{Moelans2008}. The procedure requires initializing value of interface energy $\sigma_{init}$, which has decisive role for the parameters values. For inconvenient choice of $\sigma_{init}$, some of the parameters $\gamma_{i,j}$ may get out of range of applicability of the polynomials in Table~\ref{tab_param_polynomial_coeff_myfits}. Systematic investigation revealed that the value $\sigma_{init}=(\sigma_{max}+\sigma_{min})/2$ prevents occurrence of such effects for anisotropies $\sigma_{min}/\sigma_{max}\geq 0.03$. \\
	Here we propose a simplified, yet equivalent single-step procedure. We assume constant interface width $l$ and an initializing interface energy $\sigma_{init}$, for which $\gamma=1.5$ and thus the analytic relations between the model parameters hold. As in the original iterative algorithm, the barrier height is computed
	\begin{equation} \label{eq_pardet_IWc_m}
		m = 6\frac{\sigma_{init}}{l} .
	\end{equation}
	Then, the gradient energy coefficient $\kappa_{i,j}$ can be expressed from both equations~(7) and (8), hence we can write
	\begin{equation} \label{eq_pardet_IWc_kappa}
		\kappa_{i,j}=f_{0,c}(\gamma_{i,j})ml^2 = \sigma_{i,j}l\frac{\sqrt{f_{0,c}(\gamma_{i,j})}}{g(\gamma_{i,j})} \,,
	\end{equation}
	from where we obtain
	\begin{equation}
		g(\gamma_{i,j})\sqrt{f_{0,c}(\gamma_{i,j})}=\frac{\sigma_{i,j}}{ml} = \frac{\sigma_{i,j}}{6\sigma_{init}} \,.
	\end{equation}
	In the last equation, $m$ was substituted from~\eqref{eq_pardet_IWc_m}. We denote $G(\gamma_{i,j})=g(\gamma_{i,j})\sqrt{f_{0,c}(\gamma_{i,j})}$. This function $\gamma\rightarrow G$ can be evaluated from the data tabulated in~\cite{Ravash2017} and its inverse function $G \rightarrow \frac{1}{\gamma}$ was fitted by 6-th order polynomial producing the coefficients in the last row of Table~\ref{tab_param_polynomial_coeff_myfits}. Apparently, evaluation of this polynomial in points $\sigma_{i,j}/6\sigma_{init}$ produces the searched $1/\gamma_{i,j}$. The gradient energy coefficients are then best computed as in \ref{eq_pardet_IWc_kappa}, where $m$ was substituted from \ref{eq_pardet_IWc_m}:
	\begin{equation}
		\kappa_{i,j}=6f_{0,c}(\gamma_{i,j})\sigma_{init}l \,.
	\end{equation}
	Here, the fitted polynomial $\frac{1}{\gamma} \rightarrow \sqrt{f_{0,c}}$ is used to obtain the values of $f_{0,c}(\gamma_{i,j})$.
	
	Similarly like in IWcG model, there are conditions limiting suitable values of $\sigma_{init}$
	\begin{equation} \label{eq_paramdet_IWc_siginit_ineq}
		\frac{\sigma_{max}}{6G(40)} \leq \sigma_{init} \leq \frac{\sigma_{min}}{6G(0.52)} \,,
	\end{equation}
	where $G(0.52)\approx0.0069, G(40)\approx0.4065$. Also must be satisfied
	\begin{equation}
		\frac{\sigma_{min}}{\sigma_{max}} \geq\frac{G(0.52)}{G(40)} = 0.017 \,. 
	\end{equation}
	The main advantage of the proposed algorithm is that no iterations are needed for parameters determination. In the original algorithm the number of iteration steps is uncertain (tens of repetitions are not exceptional though) and sometimes the results may not converge due to reasons tedious to debug. Obtaining each parameter in a single step thus provides better control over the process and simplifies the implementation. Moreover, the permissible values of $\sigma_{init}$ and the maximal anisotropy $\sigma_{min}/\sigma_{max}$ are clearly defined. \\
	Also, it should be noted that when $\sigma_{init}$ is chosen close to the upper limit in the inequality~\ref{eq_paramdet_IWc_siginit_ineq}, the determined $\gamma_{i,j}$s are rather closer to value 0.52 (corresponding to long-tailed interfaces). When $\sigma_{init}$ is near the bottom limit, the values of $\gamma_{max}$ are close to the value 40 and $\gamma_{min}$ are much larger than in the previous case. Near the limiting anisotropy $\sigma_{min}/\sigma_{max}\approx0.017$, the values always are $\gamma_{max}\approx40$ and $\gamma_{min}\approx0.52$. It is especially near the limits of admissible $\sigma_{init}$, where the original iterative parameters assignment strategy is not always reliable when compared to the proposed one.
	
	\subsection{IWvK}
	When $\gamma=const=1.5$, the analytic relations between the model parameters hold (i.e. equations~(8)-(10)). Then, the minimal interface width $l_{min}$ is controlled when
	\begin{equation}
		m = 6\frac{\sigma_{min}}{l_{min}}
	\end{equation}
	and
	\begin{equation}
		\kappa_{i,j}=\frac{9}{2}\frac{\sigma_{i,j}^2}{m} \,.
	\end{equation}
	Interface width of an interface $i$-$j$ can then be computed as
	\begin{equation}
		l_{i,j} = 6\frac{\sigma_{i,j}}{m} \,.
	\end{equation}
	
	\section{Properties of the used anisotropy function}\label{sec_appendix_anisofun_props}
	Let $f(\theta)$ be a smooth anisotropy function in 2D (see e.g.~\cite{Kobayashi2001} for more details). The Wulff shape $\bm{w}$ in 2D is a curve minimizing the interface energy of a crystal with constant area. It can be parameterized by the interface normal angle $\theta$, i.e.  $\bm{w}(\theta)=[w_x(\theta),w_y(\theta)]^\mathrm{T}$, giving \cite{Burton1951,Kobayashi2001,Eggleston2001}
	\begin{align} 
		w_x(\theta) &= R_W[f(\theta)\cos(\theta) - f'(\theta)\sin(\theta)] \label{eq_wulff_parametrically_x}\\
		w_y(\theta) &= R_W[f(\theta)\sin(\theta) + f'(\theta)\cos(\theta)] \label{eq_wulff_parametrically_y}\,,
	\end{align}
	where $R_W$ is the radius of the Wulff shape (arbitrary). The radius of curvature $\varrho(\theta)$ of this Wulff shape is locally
	\begin{equation} \label{eq_Wulff_radius_curvature}
		\varrho(\theta) = R_W[f''(\theta) + f(\theta)] \,,
	\end{equation}
	which may technically get negative for some $\theta$. Because that does not have any geometric interpretation, the Wulff shape does not contain such oriented segments and the angles $\theta_m$ fulfilling $\varrho(\theta_m)\leq0$ are called missing or forbidden. \\
	In the following, we assume the anisotropy function 
	\begin{equation}
		f(\theta) = 1 + \delta\cos(n\theta)
	\end{equation}
	with strength of anisotropy $0\leq\delta<1$ and $n$ an order of symmetry (integer). The interface energy is $\sigma = \sigma_0f(\theta)$. In this case, the missing angles occur for sufficiently strong anisotropy, i.e. when
	\begin{equation} \label{eq_cond_stronganiso}
		\delta\geq\frac{1}{n^2-1} \,,
	\end{equation}
	and span over an interval around every maximum of anisotorpy function $f(\theta)$. The stronger the anisotropy, the wider the interval of missing angles. Within $-\frac{\pi}{n}\geq\theta\leq\frac{\pi}{n}$, the border points of that interval are found by solving~\cite{Eggleston2001}
	\begin{equation}
		\frac{\mathrm{d}}{\mathrm{d}\theta}\frac{\cos(\theta)}{f(\theta)} = 0 \,.
	\end{equation}
	As the authors are unaware of analytic solution to this equation, it was solved numerically. The other (n-1) segments on the Wulff shape are periodic, hence all the missing angles can be obtained by rotation of the above one by multiples of $2\pi/n$.
	
	The strong-anisotropy Wulff shape is not smooth, it contains corners (i.e. jumps in local normal orientation). When plotting such Wulff shape, the missing angles $\theta_m$ must be excluded from the $\theta$s parameterizing $\bm{w}(\theta)$, otherwise "ears" behind the corners are plotted as well. 
	
	Equation~\eqref{eq_cond_stronganiso} can be used to define normalized strength of anisotropy $\Omega = \delta(n^2-1)$, which straightforwardly distinguishes between weak anisotropy $\Omega<1$ (smooth Wulff shape) and strong anisotropy $\Omega\geq1$ (the cornered one).
	
	For a given radius $R_W$, the area of a Wulff shape $A_W$ differs from the area of a circle $A=\pi R_W^2$ by the anisotropy-dependent factor $C_W(\Omega,n) = A_W/A$. For weak anisotropy $\Omega<1$ there is an analytic expression 
	\begin{equation}
		C_W = 1-\frac{\Omega^2}{n^2-1} \,,
	\end{equation}
	but there is none for strong anisotropy $\Omega>1$. The factor $C_W(\Omega,n)$ was obtained numerically for $n=4$ in the range $0\leq\Omega\leq7.5$ (the Wulff shape was plotted, its area numerically computed and divided by $\pi R^2$). There, the $\Omega$-dependence was fitted very well  by a polynomial of 4-th order $C_W(\Omega,4)= \sum_{i=0}^4a_i\Omega^{N-i}$ with $a_0=-0.00032 ,a_1=0.00639 ,a_2=-0.04219 , a_3=0.00034 , a_4=1.00000$. 
	
	\section{Shrinkage rate of a Wulff shape} \label{sec_appendix_wulff_shrrate}
	The normal velocity of anisotropic curvature-driven interface in 2D is~\cite{Abdeljawad2018}
	\begin{equation}
		v_n(\theta,t)=-\frac{\mu}{\varrho(\theta,t)}\sigma_0[f''(\theta) + f(\theta)] 
	\end{equation}
	where $\mu$ is the interface mobility and $\sigma_0[f''(\theta) + f(\theta)]$ is interface stiffness. When it is the
	Wulff shape shrinking, the radius of curvature $\varrho(\theta,t)$ can be expressed as in equation~\eqref{eq_Wulff_radius_curvature} and the inclination-dependent part of the interface stiffness is cancelled out. The normal velocity is then constant along the perimeter
	\begin{equation} \label{eq_wulff_norm_velocity}
		v_n(t) = - \frac{\mu \sigma_0}{R_W(t)} \,.
	\end{equation}
	The sign is such that the interface moves towards the center of curvature.
	
	Because the area of a Wulff shape is $A_W=\pi R_W^2C_W(\Omega,n)$, the shrinkage rate of a Wulff shape is
	\begin{equation}\label{eq_wulff_shrrate_deriv}
		\begin{split}
			\frac{\mathrm{d}A_W}{\mathrm{d}t} &= \frac{\mathrm{d}}{\mathrm{d}t}[\pi R_W^2(t)C_W(\Omega,n)]    \\
			&= 2\pi C_W(\Omega,n) R_W(t) \frac{\mathrm{d}R_W}{\mathrm{d}t} \,.
		\end{split}
	\end{equation}
	Because  the normal velocity is constant along the perimeter, the point $R_{min}$ from equation~(37) moves also by $v_n(t)$. Importantly, the normal of interface in $R_{min}$ is aligned with the radial direction of the polar coordinates, hence we can write $v_n=\mathrm{d}R_{min}/\mathrm{d}t$. From equation~(37) then must be
	\begin{equation}
		\frac{\mathrm{d}R_{W}}{\mathrm{d}t}=\frac{1}{1-\delta}\frac{\mathrm{d}R_{min}}{\mathrm{d}t}=\frac{1}{1-\delta}v_n(t) \,,
	\end{equation}
	which together with~\ref{eq_wulff_norm_velocity} implies in \ref{eq_wulff_shrrate_deriv}
	\begin{equation}
		\frac{\mathrm{d}A_W}{\mathrm{d}t} = 2\pi\mu\sigma_0  \frac{C_W(\Omega,n)}{1-\delta} 
	\end{equation}
	
	Taylor and Cahn~\cite{Taylor1998} proved that the shrinkage rate for the equilibrium shape must be a constant, which is met by the final expression.
	
	
	\section{Additional remarks on numeric implementation} \label{sec_appendix_numerics}
	In finite difference method, the ratio of the time step to the grid spacing determines the numeric stability of the simulation. Specifically, the so-called Courant number $C$ must be $0<C<0.5$, where in this model
	\begin{equation}\label{eq_Courant_nr}
		C = \mathrm{min}[\kappa(\bm{r}) L]\Delta t/(\Delta x)^2  \,.
	\end{equation}
	In practice, the time step was computed using equation~\eqref{eq_Courant_nr} assuming the value of $C\leq0.3$. For stronger anisotropies $C$ had to be reduced farther. When $0.129\leq \sigma_{min}/\sigma_{max}\leq0.2$, the value $C=0.12$ was used, when $0.2 < \sigma_{min}/\sigma_{max}\leq0.3$, $C=0.17$ and for $\sigma_{min}/\sigma_{max} > 0.3$ the used Courant number was $C=0.3$. The latter value of Courant number was also applied in simulations with inclination-dependent interface energy and smooth Wulff shapes ($\Omega\leq1$). For cornered Wulff shapes ($\Omega>1$), the following dependence was obtained from visual inspection of the interface normal inclination field $\theta(\bm{r})$ defined in equation~(18) ($\theta(\bm{r})$ was sensitive to the numeric instabilities) during the simulation 
	\begin{equation} \label{eq_Courant_nr_wulff}
		C(\Omega) = \frac{1}{a\Omega + b} \,.
	\end{equation}
	The Table~\ref{tab_Courant_nr_wulff} contains the values of $a,b$ in the respective model modifications.
	
\begin{figure}[]
		\centering
		\includegraphics[width=0.345\textwidth]{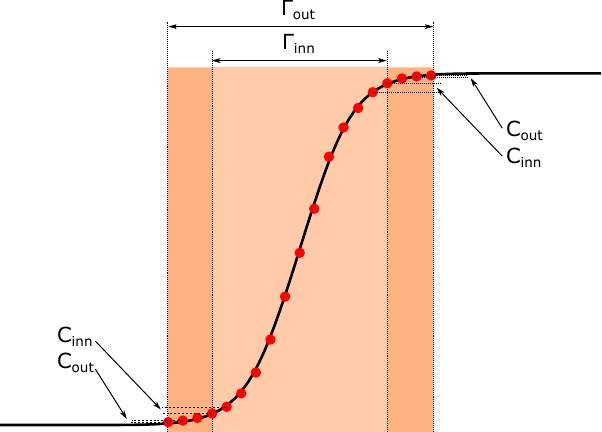}
		\caption{Illustration of the near-interface regions $\Gamma_{out},\Gamma_{inn}$, defined by absolute values of difference in phase fields between neighbouring points.}
		\label{fig_sketch_near_interf_regions}
	\end{figure}

	\begin{table}[]
		\centering
		\caption{Parameters from equation~\eqref{eq_Courant_nr_wulff} describing anisotropic dependence of the Courant number $C$ in simulations with inclination-dependent interface energy and strong anisotropy $\Omega>1$.}
		\label{tab_Courant_nr_wulff}
		\begin{tabular}{c|c|c|c}
			& IWc & IWvG & IWvK \\ \hline
			a & 1.38 & 1.43 & 0.98\\
			b& 1.95 & 1.91 & 2.35
		\end{tabular}
	\end{table}
	The following 9-point stencil was used for computation of the laplacian in grid point $(k,l)$
	\begin{equation}
		\begin{split}
			(\nabla^2\eta)_{k,l} &=  \frac{1}{6h^2}[4(\eta_{k-1,l}+ \eta_{k+1,l} \eta_{k,l-1}+ \eta_{k,l+1}) \\
			&+ \eta_{k+1,l+1} +\eta_{k-1,l-1}+\eta_{k-1,l+1}+\eta_{k+1,l-1} \\
			&-20\eta_{k,l}]  \,.
		\end{split}
	\end{equation}
	When the interface energy is inclination-dependent, it can be seen in equations~(23),~(24) of the main text that some driving force terms are proportional to $1/|\nabla\eta_i-\nabla\eta_j|$. In fact, some are even proportional to $1/|\nabla\eta_i-\nabla\eta_j|^3$ when worked out analytically. Far enough from the interface, the magnitude $|\nabla\eta_i-\nabla\eta_j|$ inevitably gets smaller than the numeric resolution, which causes severe numeric instabilities due to locally very large driving forces. For this reason, all the driving force terms from equation~(21) were only computed in a region close enough to the interface. Two such regions $\Gamma_{out},\Gamma_{inn}$ (outer and inner) were selected based on local value of $|\nabla\eta_i-\nabla\eta_j|$ (see Figure~\ref{fig_sketch_near_interf_regions}). The region $\Gamma_{inn}$ was a subset of $\Gamma_{out}$. The interface inclination field $\theta_{i,j}$ was computed in $\Gamma_{out}$, where $|\nabla\eta_i-\nabla\eta_j|^3(\Delta x)^3>C_{out}^3$ and the driving force terms $\nabla\cdot\partial f/\partial(\nabla\eta_p)$ were computed in region $\Gamma_{inn}$ where $|\nabla\eta_i-\nabla\eta_j|^3(\Delta x)^3>C_{inn}^3$. The constants $C_{out}=10^{-6}$ and $C_{inn}=0.005$ have the meaning of minimal local change in value of phase field which is acceptable to the respective near-interface region (this is exactly true for $\gamma=1.5$ and approximate otherwise). Note, that the driving force from homogeneous free energy density (i.e. $\partial f/\partial \eta_p$) and the laplacian term $\kappa(\bm{r})\nabla^2\eta_p$ were computed and applied in the whole simulation domain.
	
	In the Wulff shape simulations, the accuracy of match to the analytical shape with IWvG phase field contour turned out to be sensitive to alignment of the shape within the numerical grid. For strong anisotorpies $\Omega>1$, when the corner laid exactly on some grid line, the IWvG contours were slightly deformed, making the model perform seemingly worse than IWc and IWvK (which were not affected by this). This problem was solved by shifting the center of the initial shape by half a pixel in both $x$ and $y$ direction. No corner than laid on any grid line and that solved the problem. Rotating the Wulff shape relative to the grid had similar effect. All the results in the main text were simulated with the shifted initial shape. This is not likely to play significant role in grain growth simulations as the corner motion along grid line is very unlikely with arbitrary grain structure and local grain orientation.
	
	%
	